\documentclass[prl,twocolumn,floatfix,altaffilletter,superscriptaddress,preprintnumbers,tightenlines, ,showpacs, showkeys]{revtex4-1}

\usepackage[utf8]{inputenc}
\usepackage[colorlinks=true,citecolor=blue,linkcolor=blue]{hyperref}
\usepackage[normalem]{ulem}
\usepackage{amsmath,amssymb}
\usepackage{epsfig}
\usepackage{graphicx}               % Standard graphics package
\usepackage{url} 
\usepackage{color}
\usepackage{slashed}
\usepackage{multirow}
\usepackage{placeins}
\usepackage[dvipsnames]{xcolor}
\usepackage{epstopdf}
\usepackage{soul} 
\usepackage{tikz}
\usepackage{mathtools}
\usepackage{xcolor}

\newcommand{\beq}{\begin{equation}}
\newcommand{\eeq}{\end{equation}} 
\newcommand{\bea}{\begin{eqnarray}}
\newcommand{\eea}{\end{eqnarray}}
\newcommand{\ba}{\begin{array}}
\newcommand{\ea}{\end{array}}

\def\m1{M_1}
\def\m2{M_2}
\def\m3{M_3}

\def\ch10{\tilde \chi^0_1}

\def\gev{\,{\rm GeV}}

\def\to{\rightarrow}

\newcommand{\lsim}{\mathrel{\mathop{\kern 0pt \rlap
  {\raise.2ex\hbox{$<$}}}
  \lower.9ex\hbox{\kern-.190em $\sim$}}} 
\newcommand{\gsim}{\mathrel{\mathop{\kern 0pt \rlap
  {\raise.2ex\hbox{$>$}}}
  \lower.9ex\hbox{\kern-.190em $\sim$}}}

\definecolor{pink}{RGB}{255,105,180}

\begin{document}
%========================================

\title{
Enhancing long-lived particles searches at the LHC with precision timing information
}

\author{Jia Liu}
\affiliation{Enrico Fermi Institute, University of Chicago, Chicago, IL 60637}

\author{Zhen Liu}
\affiliation{Theoretical Physics Department, Fermi National Accelerator Laboratory, Batavia, IL, 60510}
\affiliation{Maryland Center for Fundamental Physics, Department of Physics, University of Maryland, College Park, MD 20742, USA
}

\author{Lian-Tao Wang}
\affiliation{Enrico Fermi Institute, University of Chicago, Chicago, IL 60637}
\affiliation{Kavli Institute for Cosmological Physics, University of Chicago, Chicago, IL 60637}

\date{\today}

\preprint{
%     \begin{flushright}
         FERMILAB-PUB-18-173-T, EFI-18-7 
%     \end{flushright}
}
%========================================

\begin{abstract}
We explore the physics potential of using precision timing information at the LHC in searches for long-lived particles (LLPs). 
In comparison with the light Standard Model particles, the decay products of massive LLPs arrive at detectors with time delays around
nanosecond scale. We propose new strategies to take advantage of this time delay feature by using initial state radiation to timestamp 
the collision event and require at least one LLP to decay within the detector. This search strategy is
effective for a broad range of models. In addition to outlining this general approach, 
we demonstrate its effectiveness with the projected reach for two benchmark scenarios: Higgs decaying into a pair of LLPs, and pair production of long-lived neutralinos in the gauge mediated supersymmetry breaking models. Our strategy increases the sensitivity 
to the lifetime of the LLP by two orders of magnitude or more and particularly exhibits a better behavior with a linear dependence
on lifetime in the large 
lifetime region compared to traditional LLP searches. The timing information significantly reduces the Standard Model 
background and provides a powerful new dimension for LLP searches.
\end{abstract}

\maketitle

%========================================

The presence of long-lived particles can be a striking feature of many new physics models~\cite{Barbier:2004ez,Giudice:1998bp,Meade:2010ji,Arvanitaki:2012ps,ArkaniHamed:2012gw,Liu:2015bma,Chacko:2005pe, Burdman:2006tz,Kang:2008ea,Craig:2015pha, Davoli:2017swj}. 
At the same time, vast swaths of the possible parameter space of the LLP remain unexplored by LHC searches.
LHC general purpose detectors, ATLAS and CMS, provide full angular coverage and sizable volume, making them ideal for LLP searches.
However, searches for LLPs that decay within a few centimeter of the interaction point suffer from large SM backgrounds. 
LLPs produced at the LHC generically travel slower than the SM background and decay at macroscopic distances away from the interaction point. Hence, they arrive at outer particle detectors with a sizable time delay. 

In this study, we focus on a general strategy that
uses precision timing as a tool to suppress SM backgrounds and enhances sensitivity to LLPs at the LHC. 
Recently, precision timing upgrades with a timing resolution of 30 picoseconds have been proposed to reduce pile-up for the upcoming 
runs with higher luminosity, including MIP Timing Detector (MTD)~\cite{Collaboration:2296612}  by the CMS collaboration for 
the barrel and endcap region in front of the electromagnetic calorimeter, the High Granularity Timing Detector~\cite{Allaire:2018bof} 
by the ATLAS collaboration in endcap and forward region,  and similarly multiple precision timing upgrades~\cite{Aaij:2244311} by the 
LHCb collaboration.
The usage of~(less precise) timing information for long-lived particle searches has been discussed in the past and applied to a very limited class of signals~\footnote{For a more detailed discussion, see supplemental material and references therein~\cite{Chatrchyan:2013oca, Chatrchyan:2012sp, Aaboud:2016dgf,Toback:2004xd, Goncharov:2005xz, Goncharov:2005xz, Abulencia:2007ut,Aaltonen:2008dm,Aaltonen:2013har,CMS-PAS-EXO-12-035, Aad:2014gfa,Sirunyan:2017sbs}.}. In this study,
as a strategy applicable to a broad range of models, we propose the use of a generic Initial State Radiation (ISR) jet to timestamp the hard collision and require only a single LLP decay inside the detector with significant time delay. Such a strategy can greatly suppress the SM background and 
reach a sensitivity two orders of magnitude or more better than traditional searches in a much larger parameter space~\cite{Aad:2015uaa, CMS:2014wda,Coccaro:2016lnz, Liu:2015bma}. 

With a general triggering and search strategy that can capture most LLP decays, 
we show a striking improvement in sensitivity and coverage for LLPs. In addition to the MTD at CMS, we also consider a hypothetical timing layer on the outside of the ATLAS Muon Spectrometer (MS) as an estimate of the best achievable reach of our proposal for LLPs with long lifetimes.
\footnote{For more details of our study, including the signal benchmark considerations, the time delay distributions at ATLAS MS, QCD background explanations, and trigger discussion, see supplemental material and references therein~\cite{Andersson:1983ia, Aad:2011he, jettimingtalk,ATLAS-CONF-2014-041,Shrock:1978ft,Gallas:1994xp,Ellis:2006vu, Ellis:2006oaa, Banerjee:2017hmw,Aad:2013txa,Gershtein:2017tsv}.}
\\

\noindent {\bf \textit {Basics of timing.}}---
While particle identification and kinematic reconstruction are highly developed,  
usage of timing information has so far been limited since prompt signatures are often assumed. 
Such an assumption could miss a crucial potential signature of an LLP, a significant time delay. 
Here we outline a general BSM signal search strategy that uses the timing information
and the corresponding background consideration.
\begin{figure}[htb]
    \centering
    \includegraphics[width=1.0\columnwidth]{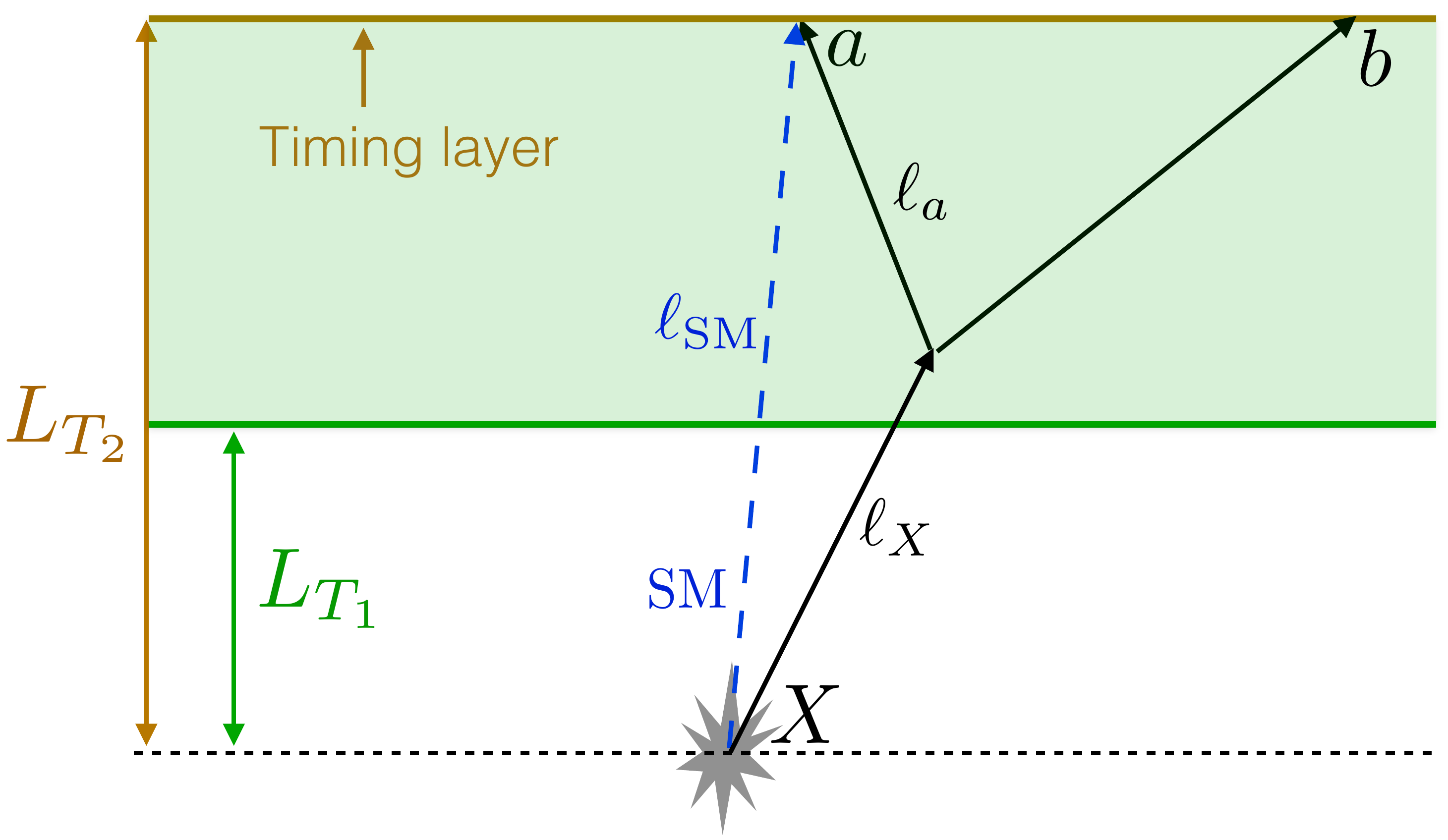} 
    \caption{An event topology with an LLP $X$ decaying into two light SM particles $a$ and $b$. A timing 
    layer, at a transverse distance $L_{T_2}$ away from the beam axis (horizontal gray dotted line), is placed at the end of the detector volume 
    (shaded region). The trajectory of a reference SM background particle is also shown (blue dashed line).
    The gray polygon indicates the primary vertex.
    }
    \label{fig:drawing}
\end{figure}
A typical signal event of LLP is shown in Fig.~\ref{fig:drawing}. 
An LLP, denoted as $X$, travels a distance $\ell_X$ into a detector volume and decays into two light SM particles 
$a$ and $b$, which then reach timing detector at a transverse distance $L_{T_2}$ away from the beam axis. Typically, the SM particles travel at velocities close to the speed of light.
For simplicity, we consider neutral LLP signals where background from charged particles can be vetoed using 
particle identification and isolation.
The decay products of X arrive at the timing layer with a time delay
\beq
 \Delta t^i_{\rm delay} = \frac{\ell_X}{\beta_X} + \frac{\ell_i}{\beta_i} - \frac{\ell_{\rm SM}}{\beta_{\rm SM}},
\label{eq:delaysimple}
\eeq
for $i$th decay products from $X$ and $\beta_i \simeq \beta_{\rm SM} \simeq 1 $. 
It is necessary to have prompt particles from production or decay, or ISR, which arrives at timing layer with 
the speed of light, to derive the time of the hard collision at the primary vertex (to ``timestamp'' the hard collision).

In Fig.~\ref{fig:DeltaTdistribution}, we show typical time delay $\Delta t$ distribution for CMS MTD for benchmark 
signals and the backgrounds.
The two benchmark signals considered here are the glueballs from Higgs boson 
decays, and the neutralino and chargino pair production in the Gauge Mediated SUSY Breaking (GMSB) scenario \cite{Giudice:1998bp,Meade:2010ji}. 
Both the glueballs and lightest neutralino proper lifetimes are set to have $c \tau= 10$~m.
The 10 GeV glueballs 
have larger average boost comparing to the 50 GeV glueballs, 
and hence have a 
sizable fraction of the signals with delays less than 1 ns.
The GMSB signal is not boosted and hence significantly delayed compared to the backgrounds, with more than 70\% of the signal having $\Delta t > 1$~ns. \\

\begin{figure}[t]
    \centering
    \includegraphics[width=1.0\columnwidth]{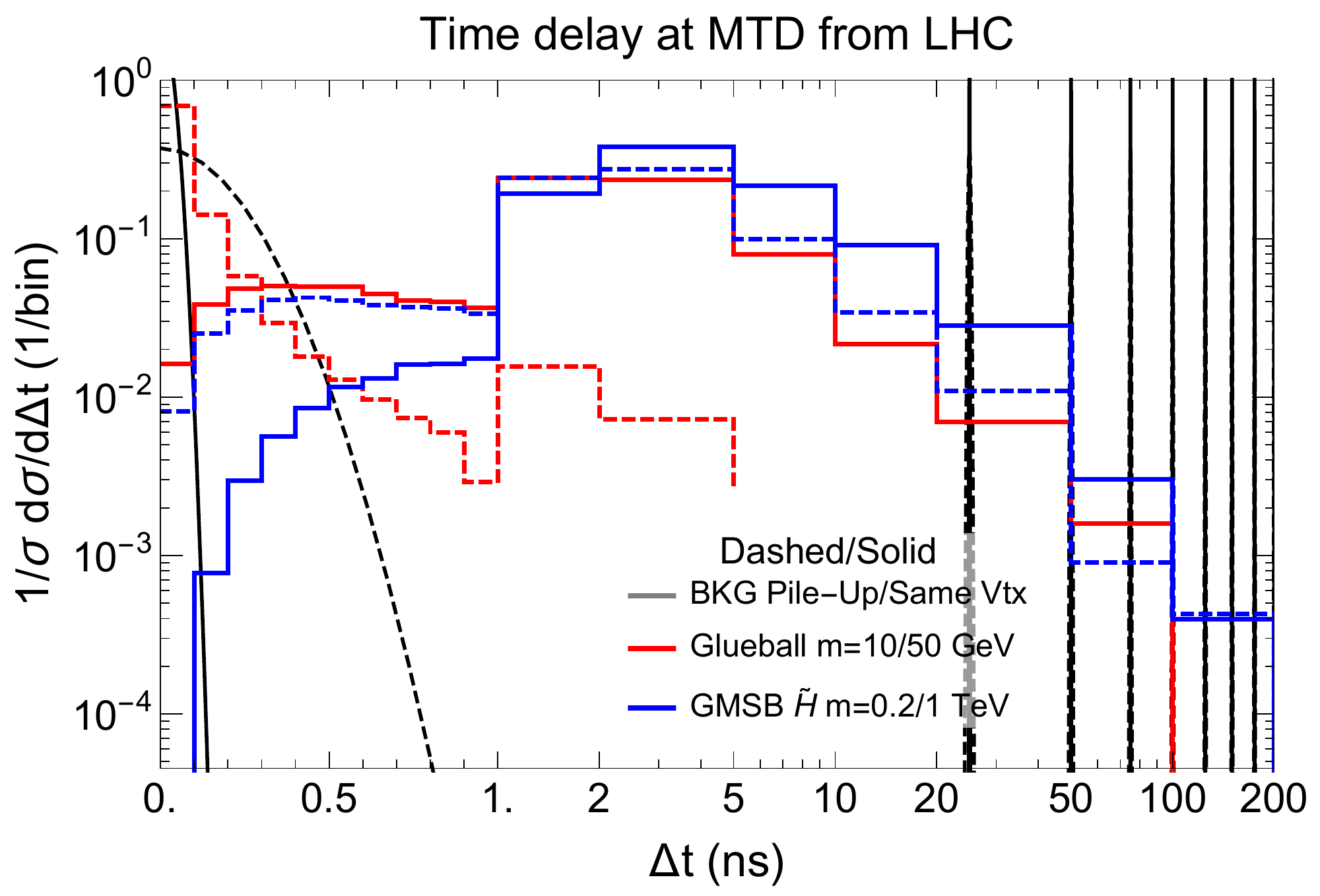} 
    \caption{The differential $\Delta t$ distribution for typical signals and backgrounds at 13 TeV LHC. The plot is normalized to the fraction of events per bin with varying bin sizes, in linear ($\Delta t < 1$ ns) and logarithmic scale ($> 1$ ns) respectively.
   Two representative signal models are shown with different masses.  The LLP proper lifetime is set to 10~m, and the distribution only counts events decayed within $[L_{T_1},~L_{T_2}]$ of [0.2,~1.17]~m in the transverse direction, following the geometry of CMS MTD in the barrel region. For the background distribution shown in gray curves, we assume bunch spacing of 25~ns. The solid and dashed gray curves represent backgrounds from the same hard collision vertex and hence with a precision timing uncertainty of $\delta_t^{\rm PT}=30$~ps and from the pile-up with a spread of $\delta_t=190$~ps, respectively.
}
    \label{fig:DeltaTdistribution}
\end{figure}

\noindent {\bf \textit {Search strategy.}}---
We consider events with at least one ISR jet to timestamp the PV and one delayed SM object coming from the LLP decay. We propose two searches using the time delay information:
\begin{center}
	\scalebox{0.95}{
    \begin{tabular}{|c|c|c|c|c|c|c|c|}
        \hline
        & $L_{T_2}$ & $L_{T_1}$ & Trigger & $\epsilon_{\rm trig}$ & $\epsilon_{\rm sig}$ & $\epsilon_{\rm fake}^{j}$ & Ref. \\ \hline
        MTD & 1.17 m & 0.2 m & DelayJet & 0.5 & 0.5 & $10^{-3}$ & \cite{Collaboration:2296612} 
        \\ \hline
        MS & 10.6 m & 4.2 m & MS RoI & 0.25, 0.5 & 0.25 & $5\times 10^{-9}$ & \cite{Aad:2015uaa}
        \\
        \hline
    \end{tabular}}
\end{center}
The size of the detector volume is described by transverse distance to the beam pipe from $L_{T_1}$ to $L_{T_2}$, where $L_{T_2}$ is the timing layer location and $L_{T_1}$ is the minimal displacement requirement for a analysis. For both searches, we assume a similar timing resolution of 30~ps.
For the MS search, because of the larger time delay and much less background due to ``shielding'' by inner detectors, 
a time resolution of 0.2 - 2 ns could achieve a similar physics reach. The $\epsilon_{\rm trig}$, 
$\epsilon_{\rm sig}$ and $\epsilon_{\rm fake}^j$ are the efficiencies for trigger, signal selection and a QCD jet faking the delayed jet signal with $p_T>30$~GeV in MTD and MS searches, respectively. 

For the MTD search, we assume a new trigger strategy dubbed ``DelayJet'' using precision timing information at CMS.
This can be realized by putting a minimal time delay cut when comparing the prompt timestamping jet~(with $p_T > 30$~GeV) with the arrival time of {\it another} jet~(with $p_T > 30$~GeV) at the timing layer. In supplemental material section (d), we describe some of the recent effort by the experimental collaboration to implement this in the triggering upgrade.

The MTD signal, after requiring $L_{T_1}$ of 0.2 m, will not 
have good tracks associated with it. Hence, the major SM background is from trackless jets.
The jet fake rate of $\epsilon_{\rm fake}^{\rm j, MTD} = 10^{-3}$ is estimated
using {\tt Pythia}~\cite{Sjostrand:2007gs} by simulating the jets with minimal $p_T$ of 30~GeV and study the anti-kt jets with $R=0.4$, where all charged constituent hadrons are too soft ($p_T<1~\gev$). 
For comparison with other studies, see supplemental material section (c).

For the MS search, we use the MS Region 
of Interest (MS RoI) trigger from a very similar search~\cite{Aaboud:2017iio} as a reference, with an efficiency 
of $\epsilon_{\rm trig}=0.25$ and 0.5 for the two benchmark BSM signals, and a signal selection efficiency of 
$\epsilon_{\rm sig}=0.25$. The backgrounds are mainly from the punch-through jets, and its  fake efficiency can be inferred 
to be $\epsilon_{\rm fake}^{\rm j, MS} = 5.2 \times 10^{-9}$, normalized to 1300 fake MS barrel events at 8 TeV~\cite{Aaboud:2017iio}, see details in supplemental material section (c). 
\\

\noindent {\bf \textit {Background consideration.}}---
The main sources of the SM background faking the delayed and displaced signal are from jets or similar hadronic activities. 
The origin of background can be classified into same-vertex (SV) hard collision and pile-up (PU). 
For this study, we assume the time-spread distributions follow a Gaussian distribution. 
\beq
\frac {d \mathcal{P}(\Delta t)} {d\Delta t} = \frac 1 {\sqrt{2}\delta_t} E^{-\frac {\Delta t^2} {2\delta_t^2}},
\eeq
where the time spreads $\delta_t$ differ for different sources of backgrounds.
The validity of these description 
should be scrutinized by experimental measurement, e.g. from Zero-Bias events. From
Refs.~\cite{Collaboration:2296612,Chatrchyan:2013oca, Chatrchyan:2012sp,Aad:2014gfa}, 
the Gaussian description is appropriate up to 
probability of $10^{-4}$ to $10^{-6}$ level. Even in the case the Gaussian fails at the tail,
a suppression power of $10^{-5}$ is already enough for MS. For MTD, one can require two time delayed objects 
to double the Gaussian suppression. Since the time delay is dominated by slow movement of $X$, the two jets 
from $X$ decay satisfy this requirement easily. 

The SV background mainly comes from QCD multi-jet production.  
At least one prompt jet is required to timestamp the event, while another trackless jet from the same 
hard collision fakes long-lived signals. The fake jet has an intrinsic time delay $\Delta t=0$.
However, it spreads out in time due to finite timing resolution, $\delta_t^{\rm PT} =30~\text{ps}$. 
At 13~TeV with $\mathcal{L}_{\rm int} = 3~\text{ab}^{-1}$, 
the estimated number of background events are% can be estimated, 
\begin{align}
&{\rm MTD}:&N_{\rm bkg}^{\rm SV} = \sigma_{\rm j} \mathcal{L}_{\rm int} {\epsilon_{\rm trig}^{\rm MTD} {\epsilon^{j, \rm{MTD}}_{\rm fake}}}  
\approx 1 \times 10^{11}  \nonumber \\
&{\rm MS}:&N_{\rm bkg}^{\rm SV} = \sigma_{\rm j} \mathcal{L}_{\rm int} {\epsilon_{\rm trig}^{\rm MS} {\epsilon^{j, \rm{MS}}_{\rm fake}}}  
\approx 4 \times 10^{5},
\label{eq:bkgQCDforHC}
\end{align}
where $\sigma_{\rm j} \simeq  1\times 10^8  $ pb is the multi-jets cross-section with two jets $p_T^j > 30$~GeV,
$\epsilon_{\rm trig}$ and $ \epsilon_{\rm fake}^j$ are the trigger and fake-rate efficiencies {\it without} using timing information.

The PU background contains two hard collisions within the same bunch crossing but do not occur at the same time. The PU background requires the coincidence of a triggered hard event and a fake signal event from pile-up
collision whose primary vertex fails to be reconstructed. At the HL-LHC, the total number of background events can be estimated,
\begin{align}
&{\rm MTD}:N_{\text{bkg}}^{\rm PU} = \sigma_{\rm j} \mathcal{L}_{\rm int} \epsilon_{\rm trig}^{\rm MTD}  \left(\bar n_{\rm PU} 
\frac{\sigma_{\rm j}} {\sigma_{\rm inc}} \epsilon_{\rm fake}^{j, {\rm MTD}} f_{\rm nt}^j \right) 
\approx 2 \times 10^7,\nonumber \\
&{\rm MS}:~~N_{\text{bkg}}^{\rm PU} = \sigma_{\rm j} \mathcal{L}_{\rm int} \epsilon_{\rm trig}^{\rm MS}  \left(\bar n_{\rm PU} 
\frac{\sigma_{\rm j}} {\sigma_{\rm inc}} \epsilon_{\rm fake}^{j, {\rm MS}} f_{\rm nt}^j \right) 
\approx 50 , \label{eq:bkgPUEcal}
\end{align}
where $\sigma_{\rm inc} = 80 ~ \text{mb}$ is the inelastic proton-proton cross-section at 13~TeV \cite{Aaboud:2016mmw}.
$\bar{n}_{\rm PU} \approx 100$ (nominally 140 or
 200~\cite{Apollinari:2116337}) is the average number of inelastic interactions per bunch crossing at HL-LHC. 
In Eq.~(\ref{eq:bkgPUEcal}), one hard collision needs to timestamp the event, while the other hard collision contains at least two jets, all of which have to be neutral to miss the primary vertex reconstruction. Otherwise, this second hard collision will leave tracks and reconstructed as another vertex in the tracker, thus get vetoed. Therefore, the background $N_{\text{bkg}}^{\rm PU}$ is suppressed by at least one additional factor of neutral jet fraction $f_{\rm nt}^j\simeq 10^{-3}$. This additional factor $f_{\rm nt}^j$, more strictly speaking, should be the probability for a multijet process whose primary vertex fails to be reconstructed and mis-assigned to the triggered vertex, which need to be estimated through full detector simulation and calibrated with data.

The key difference between the PU and SV backgrounds is that the time spread being 
determined by the beam property for the former (190 ps~\cite{Collaboration:2296612}), and by the timing resolution for the latter (30 ps~\cite{Collaboration:2296612}).  
For the MTD (MS) search, if we apply cut $\Delta t > 1$~(0.4)~ns, the total estimated events from SM 
background is 1.3 (0.86), where the SV background become completely negligible.

Backgrounds not from the hard collision have larger temporal spread, such as cosmic ray, beam halo, misconnected tracks, interaction with detector material, etc. At the same time, their properties are well measured and can be vetoed effectively. For example, for the MS search, displaced vertex reconstruction can help suppress the above backgrounds. Its efficiency has been included in $\epsilon_{\rm sig}$~\cite{Aad:2015uaa}. In another example, the non-pointing photon searches study at ATLAS \cite{Aad:2014gfa} found such backgrounds are negligible, with two photon final states which only have directional information. Ref.~\cite{Sirunyan:2017sbs} measured the stopped particle signatures and found that \textit{ the energy cut alone} can reduce the background to single digit. In comparison, our signal has more kinematic features, such as large energy deposition (more than 30 GeV) and high track multiplicities with sizable time-delay. It can be further separated from these backgrounds. The argument for MS also applies to the CMS MTD search. The search for a pair of jets from one displaced vertex \cite{CMS:2014wda} found SM QCD background to be dominant. Moreover, since MTD detector is much smaller than MS, the cosmic ray background is less problematic. Even assuming the number of SM background events to be 100, the limits in Fig. 3 are only weakened by factor of 5. \\

\noindent {\bf \textit {Augmented sensitivity on LLPs through precision Timing.}}---
Our first example, Signal A (SigA), is Higgs decaying to glueballs with subsequent decays into SM jet pairs. This occurs in model~\cite{Craig:2015pha} where the Higgs is the portal to a dark QCD sector whose lightest states are the long-lived glueballs. 
Typical energy of the glueball is set by the Higgs mass, and the time delay depends on glueball mass. 

The second example, Signal B (SigB), is the decay of the lightest neutralino in the GMSB scenario.  Its decay into SM bosons 
($Z$, h, or $\gamma$) and gravitino is suppressed by the SUSY breaking scale $\sqrt F$, and it can be naturally long-lived.
This benchmark represents the timing behavior of pair produced particles at the LHC without an intermediate resonance.

For both examples, timestamping the hard collision is achieved by using an ISR jet: 
\bea
\rm{SigA:}& \ pp\rightarrow h+j&,\ h\rightarrow X+X,\ X\rightarrow {\rm SM}   ,\\
\rm{SigB:}& \ pp\rightarrow \tilde \chi \tilde \chi+j&,\ \tilde \chi^0_1 \to h+\tilde G \to {\rm SM}+\tilde G.
\eea
For SigB, other electroweakinos $\tilde \chi$, such as charginos $\tilde \chi^\pm$ or heavier neutralino $\tilde \chi^0_2$, 
promptly decay into the lightest neutralino state $\tilde \chi^0_1$ plus soft particles. Hence, we take the inclusive Higgsino pair production cross-section for this process.

\begin{figure}[t]
    \centering
    \includegraphics[width=1.0\columnwidth]{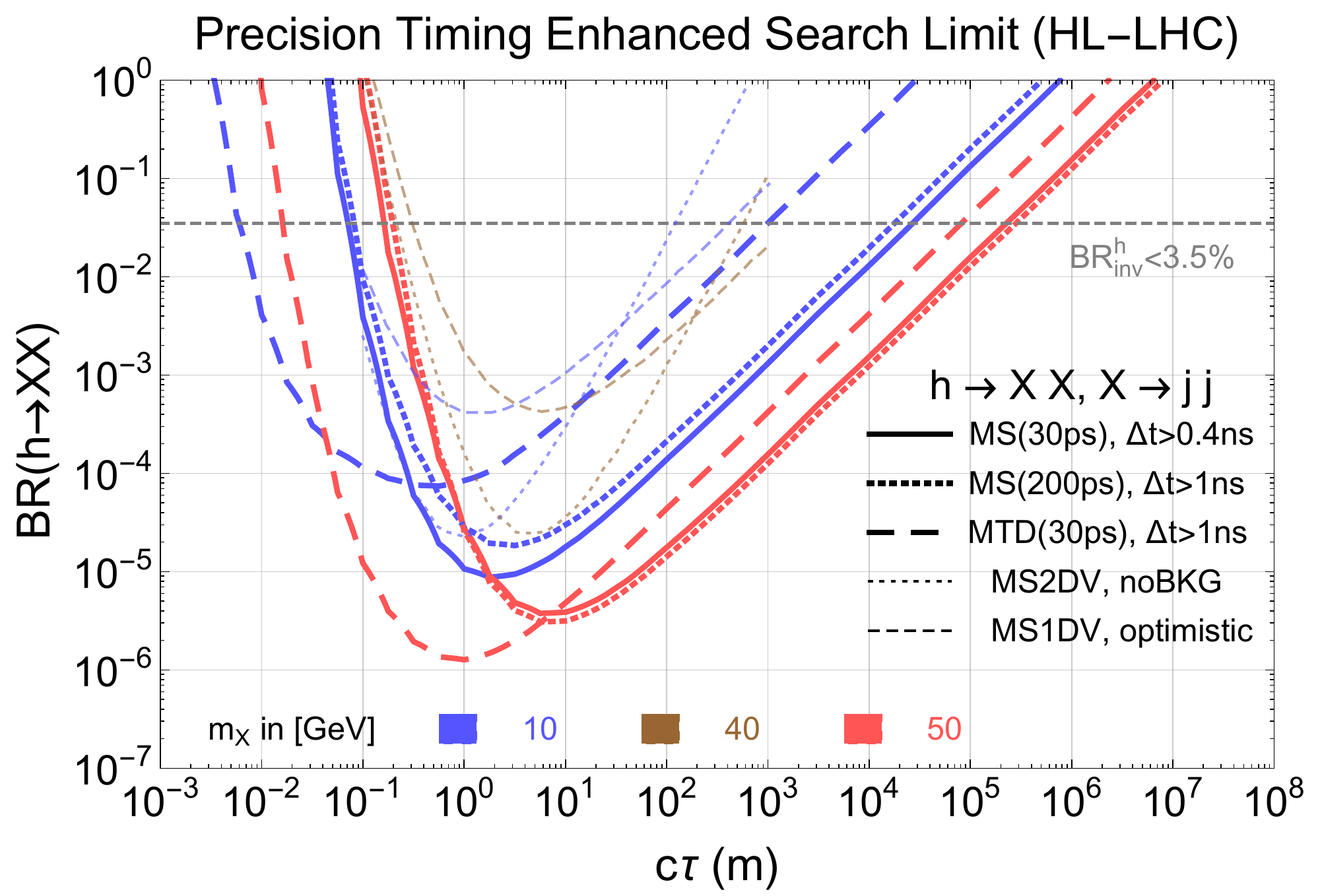} 
    \caption{The $95\%$ C.L. limit on $\text{BR}(h \to XX)$ for signal process $pp \to j h$ with subsequent decay 
    $h\to X X$ and $X \to j j$. Different colors indicate different masses of the particle $X$. 
    The thick solid and dotted (thick long-dashed) lines indicate MS (MTD) searches with different timing cuts. The numbers in parentheses are 
    the assumed timing resolutions. Other 13 TeV LHC projections \cite{Coccaro:2016lnz, Bernaciak:2014pna} are plotted in thin lines.
    }
    \label{fig:ctaulimitHiggs}
\end{figure}
To emphasize the power of timing, we rely mostly on the timing information to suppress background and make only 
minimal cuts. We only require one low $p_T$ ISR jet, with $p_T^j > 30~\text{GeV}$ and $|\eta_j| < 2.5$.
In both signal benchmarks, we require at least one LLP decays inside the detector. We generate signal events using {\tt MadGraph5}~\cite{Alwall:2014hca} at parton level and adopt the UFO model file from~\cite{Christensen:2013aua} for the GMSB simulation. 
After detailed simulations of the delayed arrival time, we derive the projected sensitivity to SigA and SigB using the 
cross-sections obtained in Ref.~\cite{Greiner:2015jha} and Refs.~\cite{Fuks:2012qx,Fuks:2013vua}, respectively.

For SigA, the $95\%$ C.L. sensitivity is shown in Fig.~\ref{fig:ctaulimitHiggs}. 
We assume $X$ decays to SM jet pairs with 100\% branching fraction.
The MTD and MS searches, with $30$ ps timing resolution, are plotted in thick dashed and solid lines, respectively. 
For MS, the best reach of $\text{BR}(h\to XX)$ is about a few $10^{-6}$ for $c \tau < 10$~m. 
It is relatively insensitive to the mass of $X$ when $m_X> 10~\gev$ because $X$ are moving slowly enough
to pass the timing cut. 
For the MS search, a less precise timing resolution 
($200$ ps) has also been considered with cut $\Delta t > 1$ ns. After the cut, the backgrounds 
from SV and PU for MS search are $0.11$ and $7.0 \times 10^{-3}$ respectively, and the SV background dominates.  
The reach for heavy $X$ is almost not affected, 
while reduced by a factor of $\sim 2$ for light X.

In Fig.~\ref{fig:ctaulimitHiggs}, we compare MTD and MS (thick lines) with 13 TeV  HL-LHC (with $3 ~\text{ab}^{-1}$ integrated luminosity) projections, two displaced vertex (DV) at 
MS using zero background assumption (thin dotted) and one DV at MS using a data-driven method with optimistic background estimation (thin dashed) from \cite{Coccaro:2016lnz}. 
The projected limits from invisible Higgs decay at 13 TeV~\cite{Bernaciak:2014pna} is also shown in Fig.~\ref{fig:ctaulimitHiggs}.

\begin{figure}[t]
    \centering
    \includegraphics[width=1.0\columnwidth]{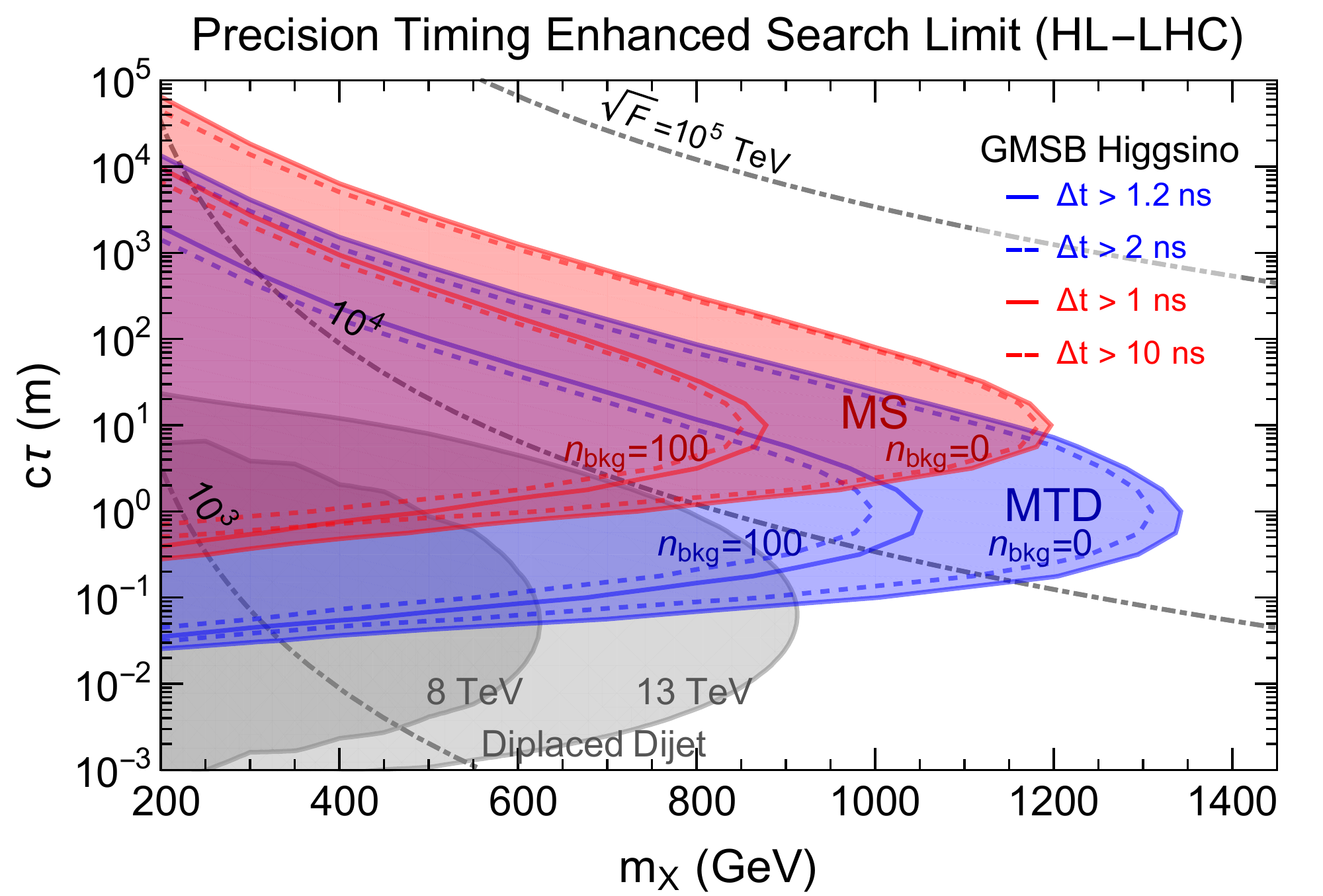} 
    \caption{The projected $95\%$ C.L. limit on the Higgsino mass--lifetime plane 
        for signal process of Higgsino pair production in association with jets, with subsequent decay of the lightest Higgsino $\tilde{\chi}^0 \to h \tilde{G}$ 
        and $h \to b b$ in GMSB scenario. We decouple other electroweakinos and have Higgsino-like chargino ${\tilde{\chi}^{\pm}}$ and neutralino ${\tilde{\chi}^0_2}$ nearly degenerate with $\tilde{\chi}^0_1$.
   }
    \label{fig:DeltaTdistributionForneu2}
\end{figure} 
For SigB, we show the projected 95\% C.L. exclusion reach in the plane of Higgsino mass $m_{\tilde \chi}$ and proper 
lifetime $c\tau$ in Fig.~\ref{fig:DeltaTdistributionForneu2}. The projected coverage of the MTD and MS searches in blue and
red shaded regions, respectively. Due to the slow motion of $\tilde{\chi}$, we show the projections with a tight (solid lines) 
and a loose (dashed lines) $\Delta t$ requirement. 
The loose selection, $\Delta t >10$~ns allows us to use the current muon timing resolution of 2~ns~\cite{Sirunyan:2018fpa} to achieve similar coverage.
Although MTD and MS searches with $\Delta t > 1$ and $0.4$ ns cuts have background event
of order 1, we also show the sensitivity reach with a sizable background of 100 at the HL-LHC. 
We observe a similar behavior for the coverage of MTD and MS searches in term of the lifetime for SigB.

Furthermore, we draw gray dashed-dotted lines for SUSY breaking scale $\sqrt F$.
To compare with existing LLP searches and their projection, 
we follow Ref.~\cite{Liu:2015bma} and quote the most sensitive CMS displaced dijet search conducted at 8 TeV~\cite{CMS:2014wda}, and show 
the projected sensitivity at 13 TeV HL-LHC assuming statistical dominance for the background. 
We can see 
timing searches almost double the reach of $m_{\tilde{ \chi }}$ with lifetime around one meter, and
extend the sensitivity to very long lifetime, up to $10^5$~m for a 200~GeV LLP. \\

\noindent{\bf\textit{Discussion.}}--We demonstrate in this letter that exploiting timing information can significantly enhance 
the sensitivities of LLP searches at the LHC. 
To emphasize the advantage of timing, we made minimal requirements on the signal, with one ISR jet
and a delayed signal. Further optimization can be developed for more dedicated searches. 
The timestamping ISR jet can be replaced by other objects, such as leptons or photons. Depending on the underlying signal 
and model parameters, one can also use prompt objects from signal production and decay. In addition, for specific searches, 
one could also optimize the selection of the signal based on the decay products of LLPs. 
Finally, we emphasize that the current LLP searches are complementary to the timing proposed in this letter. Once combined,
the current searches should in general gain better sensitivity for heavy LLP. \\

\begin{acknowledgements} \noindent {\bf\textit{Acknowledgment:}}-We would like to thank specifically to David Curtin, Simon Knapen, Si Xie and Charles Young for insightful comments. We also benefited from helpful discussions with Artur Apresyan, Matthew Cintron, Jared Evans, Henry Frisch, Joshua Isaacson, Matthew Low, Brian Shuve, Nhan Tran, Jessica Turner and Gordon Watts. 
	LTW is supported by the DOE grant DE-SC0013642. JL acknowledges support by an Oehme Fellowship. This manuscript has been authored by Fermi Research Alliance, LLC under Contract No. DE-AC02-07CH11359 with the U.S. Department of Energy, Office of Science, Office of High Energy Physics. ZL is supported in part by the NSF under Grant No. PHY1620074 and by the Maryland Center for Fundamental Physics.
\end{acknowledgements}

%\bibliographystyle{utphys}
%\bibliography{reference}

%\end{document}

%\begin{document}
%\clearpage
%\newpage
\maketitle
\onecolumngrid
\begin{center}
\vspace{0.5in}
	{\textbf{\Large Supplemental Material }}\\
		%\textbf{\large Enhancing Long-lived particles searches at the LHC with precision timing information}} \\ 
%	\vspace{0.05in}
%	{ \it \large Supplementary Material}\\ 
	\vspace{0.05in}
	%{Jia Liu, Zhen Liu and Lian-Tao Wang} \\
\end{center}
	{	In this supplemental material, we provide more information for: (a) the signal benchmark considerations, (b) the time delay at CMS MTD, 
	(c) (more) background considerations and (d) trigger discussions.} \\

%\begin{abstract}
%	In this supplemental material, we provide more information for: (a) the signal benchmark considerations, (b) the time delay at CMS MTD, 
%	(c) (more) background considerations and (d) trigger discussions.
%\end{abstract}

\twocolumngrid

\subsection{(a) the signal benchmark considerations}

\begin{figure}[ht!]
	\centering 
	\includegraphics[width=0.7\columnwidth]{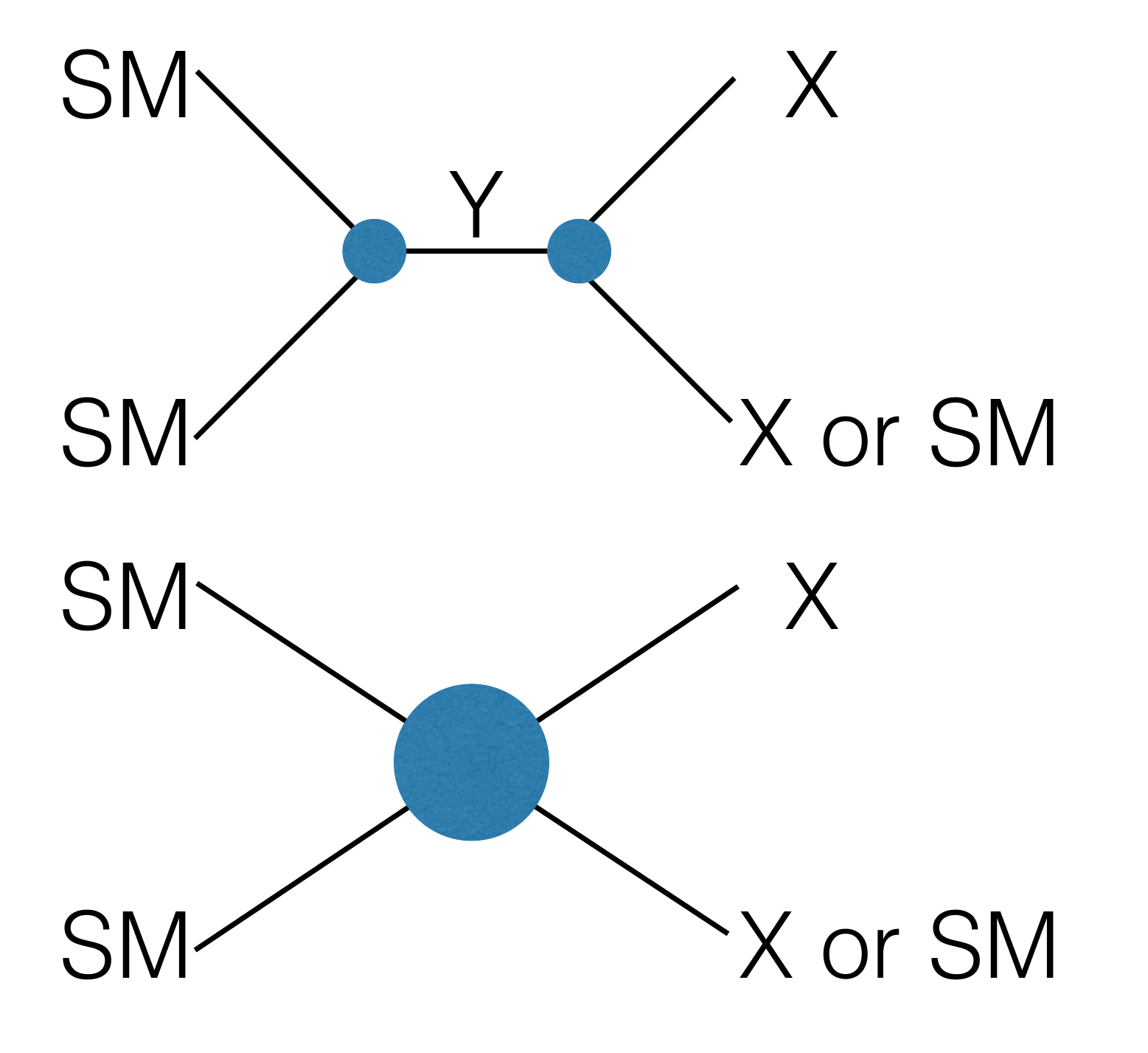} 
	\caption{Two classes of signal kinematics for LLPs.}
	\label{fig:channels}
\end{figure} 
In general, there are two classes of qualitatively different production channels for the LLPs, as shown in Fig.~\ref{fig:channels}. In the first class (upper panel), the LLP(s), denoted as $X$,  are produced through the decay of a heavier resonance ($Y$), which can contain one or more LLPs. Perhaps the most well-known model in this class is when the resonance is the Higgs boson ($Y=h$). This is highly motivated by possible connection of new physics and electroweak symmetry breaking. At the same time, the resonance can certainly be other SM particles, such as $W$, $Z$ and the top quark, and BSM particles such as $W^\prime$, $Z^\prime$, and so on. They all share some common characteristics. The rate of this process is controlled by the production rate of the resonance and the branching ratio into the LLP(s). The decay length of the LLP, $d=\gamma \beta c \tau$,  plays an important role in determining signal rate within the detector volume. Moreover, the boost $\gamma$ is also important in determining the time delay. In this class of models, the boost of the LLP is set by the mass ratio $\gamma \propto m_Y/m_X$. 

In the second class of models, shown in the lower panel of Fig.~\ref{fig:channels}, the LLP(s) can be produced directly without going through a resonance. This would be the case, for example, for heavier $X$ with SM interactions. A typical benchmark would be the production of SUSY electroweakinos. The signal of this class of models have distinct features as well. In particular, they will be produced close to the threshold, with velocity being a fraction of the speed of light. In this case, a large time delay is always expected.

This choices of SigA and SigB in the main text are chosen to capture above two representative classes of the LLP kinematics.

\subsection{(b) the time delay at ATLAS MS}

\begin{figure}[ht!]
	\centering
	\includegraphics[width=1.0\columnwidth]{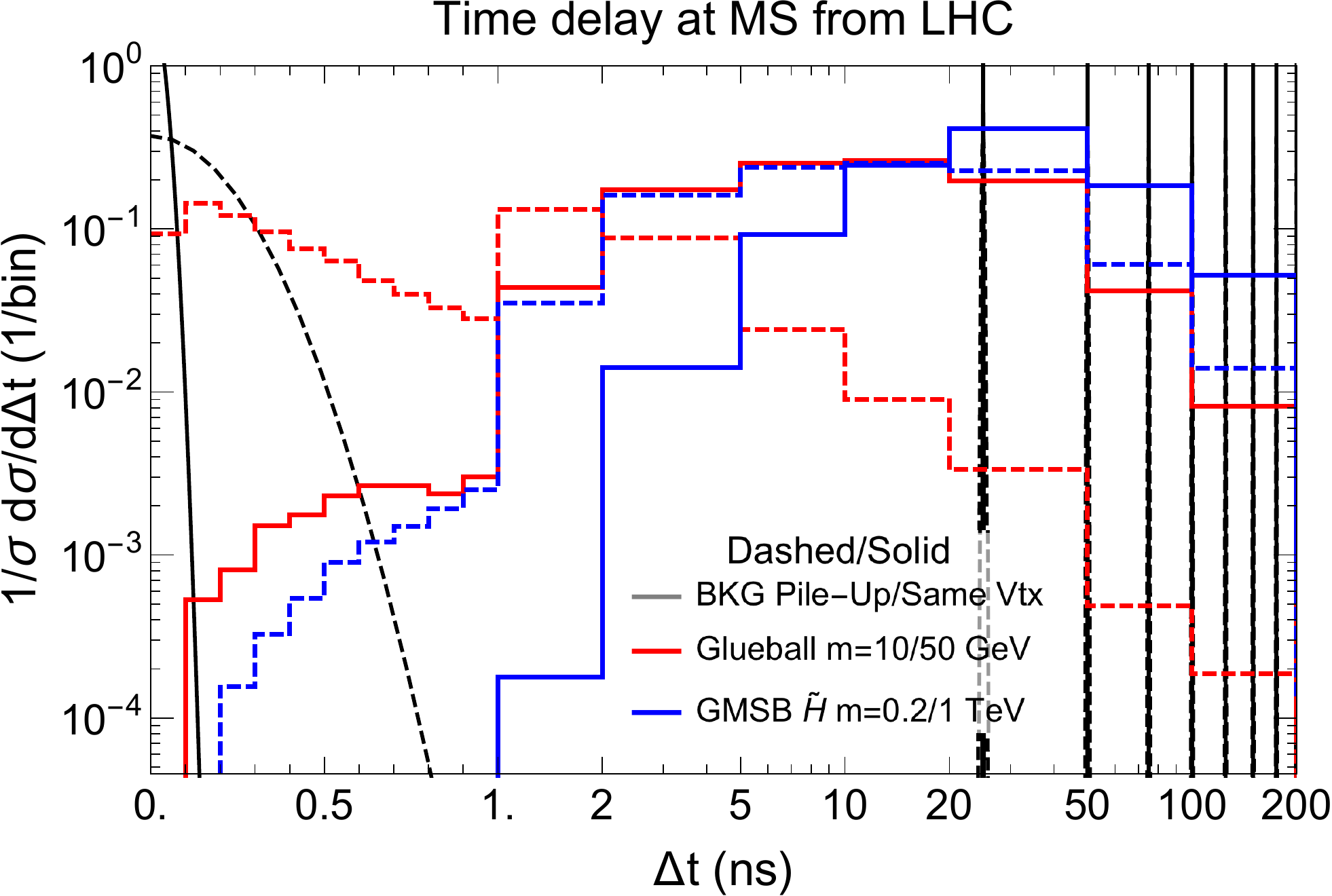} 
	\caption{The differential $\Delta t$ distribution for typical signals and backgrounds at 13 TeV LHC for ATLAS MS. 
		The legends are the same as in Fig.~2.}
	\label{fig:DeltaTdistributionEC}
\end{figure} 

Timing information has been applied to BSM searches in identifying new physics in some very limited cases.
Such examples include the time of flight parameter adopted in the heavy stable charged particle searches~\cite{Chatrchyan:2013oca, Chatrchyan:2012sp, Aaboud:2016dgf}, the time delay parameter adopted in the non-pointing photon searches at the CDF and recently at the LHC~\cite{Toback:2004xd, Goncharov:2005xz, Abulencia:2007ut,Aaltonen:2008dm,Aaltonen:2013har,CMS-PAS-EXO-12-035, Aad:2014gfa}, and (very loosely) in the stopped particle searches~\cite{Sirunyan:2017sbs}. Precision timing thus opens a new window to search for Beyond Standard Model (BSM) signals.

In Fig.~\ref{fig:DeltaTdistributionEC}, we show typical time delay $\Delta t$ for the ATLAS MS for benchmark signals and the backgrounds. In comparison with the time delay distribution for CMS MTD, shown in Fig.~2, the signal delays are enlarged by roughly an order of magnitude. After the cut $\Delta t > 1$ ns, the heavy particles in the signal are almost not affected, 
and only $10$ GeV $X$ lose some fraction of events. This fact is in good agreement with Fig.~3.

We have considered two concepts of timing layer at the LHC. 
The CMS MTD timing upgrade for HL-LHC already provides significant improvement. 
The MS system has the notable benefits of low background, a large volume for the LLP to decay and 
more substantial time delay for the LLP signal due to the longer travel distance, see Fig.~3. 
Therefore, a less precise timing resolution can still achieve similar physics goals for MS system.  
It can serve as an estimate of the best achievable sensitivity using timing information in LLP searches. 
A feasibility study on new timing layer options like this, balancing technology, design, cost, and physics goals would 
be a natural next step, given the promising results shown in this study. In summary, the precision timing 
enhanced search for LLPs is very generic and can suppress SM background significantly. The timing information 
should act as a new dimension in the future searches.

\subsection{(c) QCD background explanations}

In the ATLAS MS search, we have chosen the detector transverse length between 4.2 m and 10.6 m. However, 
the ATLAS MS displaced vertex search~\cite{Aad:2015uaa}, due to the vertex reconstruction requirement, can only effectively select signal events decaying in the 4-7~m range, reducing the derived search sensitivity with the full MS volume approximately by a factor of two. We expect that with the help of the timing layer and a relaxed vertex reconstruction requirement, the effective decay range could be extended to the full MS while maintaining the same signal efficiency.  In comparison with LLP decay in the 7-10~m range of the MS, there is no detector activities in the layers prior to that. Hence, the dominant background from punch-through jet can still be vetoed effectively.

The trajectories of charged SM particles can be curved in the magnetic field, which increase the path length in comparison with 
neutral SM particles. We use the standard jet algorithm, and define the time of jet by the first arrival objects inside the cone. With 
sufficient information in the tracker, we can even avoid using curved (low-$p_T$) tracks to define the time. In this regard, the arrival time of a jet is defined by the leading components. 
Although jets contain soft (and hence slow) particles, the majority of the 
constituent particles in a jet still travel with nearly the speed of light~\cite{Collaboration:2296612, Andersson:1983ia, Aad:2011he, jettimingtalk}.

The trackless jet fraction is measured in the validation data for the low-electromagnetism jet search at the ALTAS~\cite{ATLAS-CONF-2014-041}, and it is found to be $10^{-2}$. 
However, they also found a huge additional suppression through the energy deposition ratio between electromagnetic and hadronic calorimeters. We have calculated the trackless jet fraction using {\it Pythia} and obtained $10^{-3}$. 
Considering the suppression, our estimation is reasonable. 

The pile-up events have both time and spatial spread. Therefore, the interaction point information $z$ would also enter the 
estimation of such background. However, given that the typical spread is few cm, it can induce a time shift at most
$\approx \mathcal{O}(100) $ ps \cite{Collaboration:2296612}, typically with an addition suppression of a geometrical factor.  
Adding in quadrature, this will at most give an insignificant increase the spread in time by $\approx 60$~ps. 
One can use larger time delay cut to alleviate this effect. It has even less impact for MS search, where the pile-up background is 
already small before timing cut.

The number of SV backgrounds with 30 ps resolution are $10^{-232}$ and $10^{-35}$ for MTD and MS 
respectively, with time cut at 1 ns and 0.4 ns. For 60 ps resolution, the Gaussian suppression power decreased 
by one-quarter ($1/4$) in the exponent. The number of SV backgrounds become $10^{-51}$  and $10^{-5}$ for 
MTD and MS respectively. The SV backgrounds become negligible after timing cut compared with PU one. We note this shows for SV background, although the background seems to be big to begin with, our timing cut choice is very conservative and leave huge room for non-Gaussianity from such background.

In the future, the object reconstruction with separation not only in spatial but also
in time should help discriminate these various backgrounds. In specific searches, signal typically has additional 
feature to further suppress the background. For instance, in our case, we actually have two visible objects with 
different time delays. Taking advantage of such characteristics, we expect the background can be further suppressed.

\subsection{(d) trigger discussions}
Triggering on delayed signals concerning the primary interaction vertex could become a very interesting and important application for the general class of long-lived particle signals~\cite{Shrock:1978ft,Gallas:1994xp,Ellis:2006vu, Ellis:2006oaa, Banerjee:2017hmw}. 
Triggers with additional timing information (such as sizable delay) would complement current trigger system that focuses on very hard events, using $H_T$, $p_T$ of jets, leptons, photons, and missing $E_T$~\cite{Aad:2013txa,Gershtein:2017tsv}. A much softer threshold could be achieved with sizable time delays as an additional criterion, which would be extremely beneficial for LLP, especially for compressed BSM signal searches.

\bibliographystyle{utphys}
\bibliography{arxiv-v3}

\end{document}